\documentclass[twocolumn]{aastex61}
\received{July 1, 2016}
\revised{September 27, 2016}
\accepted{\today}
\submitjournal{ApJ}

\shorttitle{Spectroscopic of SN\,2017iuk Associated with GRB\,171205A}
\shortauthors{Wang et al.}

\begin{document}

\title{Spectroscopy of the Type Ic Supernova SN\,2017iuk Associated with Low-redshift GRB\,171205A}

\correspondingauthor{J. Wang}
\email{wj@bao.ac.cn}
\correspondingauthor{Z. P. Zhu}
\email{zipei@hust.edu.cn}
\correspondingauthor{D. Xu}
\email{dxu@nao.cas.cn}

\author{J. Wang}
\affil{Key Laboratory of Space Astronomy and Technology, National Astronomical Observatories, Chinese Academy of Sciences, Beijing
100101, China}
\affil{School of Astronomy and Space Science, University of Chinese Academy of Sciences, Beijing, China}

\author{Z. P. Zhu}
\affil{School of Physics, Huazhong University of Science and Technology, Wuhan 430074, China}
\affil{Key Laboratory of Space Astronomy and Technology, National Astronomical Observatories, Chinese Academy of Sciences, Beijing
100101, China}

\author{D. Xu}
\affil{Key Laboratory of Space Astronomy and Technology, National Astronomical Observatories, Chinese Academy of Sciences, Beijing
100101, China}

\author{L. P. Xin}
\affil{Key Laboratory of Space Astronomy and Technology, National Astronomical Observatories, Chinese Academy of Sciences, Beijing
100101, China}

\author{J. S. Deng}
\affil{Key Laboratory of Space Astronomy and Technology, National Astronomical Observatories, Chinese Academy of Sciences, Beijing
100101, China}
\affil{School of Astronomy and Space Science, University of Chinese Academy of Sciences, Beijing, China}

\author{Y. L. Qiu}
\affil{Key Laboratory of Space Astronomy and Technology, National Astronomical Observatories, Chinese Academy of Sciences, Beijing
100101, China}

\author{P. Qiu}
\affil{Key Laboratory of Optical Astronomy, National Astronomical Observatories, Chinese Academy of Sciences, Beijing
100101, China}

\author{H. J. Wang}
\affil{Key Laboratory of Optical Astronomy, National Astronomical Observatories, Chinese Academy of Sciences, Beijing
100101, China}

\author{J. B. Zhang}
\affil{Key Laboratory of Optical Astronomy, National Astronomical Observatories, Chinese Academy of Sciences, Beijing
100101, China}

\author{J. Y. Wei}
\affiliation{Key Laboratory of Space Astronomy and Technology, National Astronomical Observatories, Chinese Academy of Sciences, Beijing
100101, China}
\affiliation{School of Astronomy and Space Science, University of Chinese Academy of Sciences, Beijing, China}



\begin{abstract}
 We here report a spectroscopic monitor for the supernova SN\,2017iuk
 associated with the long-duration low-luminosity gamma-ray burst GRB\,171205A at a redshift of 0.037, which is up to now the third
 GRB-SN event away from us. 
 Our spectroscopic observations and spectral analysis allow us to identify SN\,2017iuk as a typical broad-line type Ic SN. 
 A comparison study suggests that the type-IcBL SN\,2017iuk resembles to SN\,2006aj in following aspects:  
 1) similar spectra at the nearby epochs, 2) comparable evolution of the photospheric velocity obtained from the measurements based on both 
 \ion{Si}{2}$\lambda$6355 line and spectral modeling, and 3) comparable explosion parameters. This analogy could imply
 a formation of a neutron star in the core-collapse of GRB\,171205A/SN\,2017iuk as previously suggested in GRB\,060218/SN\,2006aj. 
 The properties of the host galaxy is discussed, which suggests that GRB\,171205A/SN\,2017iuk 
 occurred in an early type (S0), high-mass, starforming galaxy with low specific SFR and solar metallicity.

\end{abstract}




\section{Introduction} \label{sec:intro}

The connection between long-duration gamma-ray bursts (LGRBs) and their associated broad-line type Ic supernovae (SNe IcBL) 
with an absolute magnitude of $M_{\mathrm{V}}\sim-19$ mag has been firmly 
established in past two decades due to the prompt spectroscopic follow-up observations (see reviews in 
Woosley \& Bloom 2006; Hjorth \& Bloom 2012; Cano et al. 2017 and references thererin). 
So far, the GRB-SN association has been identified in about 30 events with a redshift range from 
0.00867 (GRB\,980425/SN\,1998bw, e.g., Galama et al. 1998) to 1.0585 (GRB\,000911, e.g., Lazzati et al. 2001; 
Masetti et al. 2005), although a non-association was firmly identified in 
two low-$z$ cases: GRB\,060505 and GRB\,060614 by deep imaging (Fynbo et al. 2006; Della Valle et al. 2006; Gal-Yam et al. 2006). 

Study of the SNe associated with LGRBs is important for exploring the nature of death of massive stars.
Even though it is widely accepted that LGRBs originate from core-collapse of young massive stars 
(e.g., Woosley \& Bloom 2006; Hjorth \& Bloom 2012; Wang et al. 2018 and references therein),  
the compact objects formed in the core-collapse are still under hot debate. They could be 
either a stellar-mass blackhole (e.g., Woosley 1993; MacFadyen \& Woosley 1999) or a magentar 
(i.e. a rapidly rotating neutron star with extremely large magnetic field, e.g., Dai \& Lu 1998; Woosley \& Heger 2006; 
Zhang \& Dai 2010; Bucciantini et al. 2007; 
Mazzali et al. 2014; Wang et al. 2017). In fact, the SN explosion mechanism depends on the compact objects formed in the core-collapse.
For instance, two types of central engines, i.e., 1) radioactive-heating caused by the radioactive decay of nickel
and cobalt, and 2) rotation energy of a magentar, have been proposed 
for the powering of the SNe associated with LGRBs (e.g., Greiner et al. 2015; Cano et al. 2016, 2017;
Metzger et al. 2015; Bersten et al. 2016; Sukhbold et al. 2016; Wang et al. 2015; Dai et al. 2016; Iwamoto et al.
2000; Nakamura et al. 2001; Maeda et al. 2003).  

Among the identified LGRB-SN events, only two cases, i.e., GRB\,980425/SN\,1998bw and GRB\,060218/SN\,2006aj 
(e.g., Pian et al. 2006; Mazzali et al. 2006) are found to be at redshift less than 0.05 ($\sim200$ Mpc).  
Both them can be classified as low-luminosity GRBs (\it ll\rm GRBs) with an isotropic $\gamma$-ray 
luminosity of $\log L_{\gamma,\mathrm{iso}}<48.5$. In this paper, we report a spectroscopic study for the 
type-IcBL SN\,2017iuk 
associated with GRB\,171205A at a redshift of 0.037, which is up to now the third GRB-SN event away from us.

The paper is organized as follows. Section 2 summarizes the basic properties of GRB\,171205A/SN\,2017iuk. The 
spectroscopic observations and data reductions of SN\,2017iuk are presented in Section 3. Section 4 shows the 
analysis and results, A conclusion and implications are presented in Section 5.     
A $\Lambda$CDM cosmology with parameters $H_0=70\ \mathrm{km\ s^{-1}\ Mpc^{-1}}$,
$\Omega_{\mathrm{m}}=0.3$, and $\Omega_\Lambda=0.7$ is adopted throughout the paper.

\section{GRB\,171205A and Associated SN\,2017iuk} \label{sec:style}

GRB\,171205A was detected by \it Swift \rm Burst Alert Telescope (BAT) on December 05, 2017 at 
07:20:43 UT (\it Swift \rm trigger 794972,  D'Elia et al. 2017a). \it Swift \rm XRT detected 
a bright new X-ray source at 144.7 seconds after the BAT trigger (D'Elia et al. 2017b).  
The BAT on-ground analysis (Barthelmy et al. 2017) reported a duration of $T_{90} = 189.4\pm35.0$s, a 
fluence of $(3.6\pm0.3)\times10^{-6}\ \mathrm{erg\ cm^{-2}}$ in the 15-150 keV band , and 
a powerlaw spectrum with a photon index of $\Gamma=1.41\pm0.14$. 
A photon index of $\Gamma=1.717^{+0.035}_{-0.024}$ within the 0.3-10 keV band was reported by a refined XRT analysis (Kennea et al. 2017).
The burst was also detected by Konus-Wind with a well fitted a powerlaw spectrum with a
$\Gamma=2.0\pm0.14$ in the 20-1500 keV band (Frederiks et al. 2017).   
The reported fluence of $(7.8\pm1.6)\times10^{-6}\ \mathrm{erg\ cm^{-2}}$ corresponds to an
isotropic energy of $E_{\mathrm{iso}}\sim6.6\times10^{49}\ \mathrm{erg}$ in the 1-10000keV band, and to
a $\log L_{\gamma,\mathrm{iso}}=47.5$, which allows us to classify GRB\,171205A as a \it ll\rm GRB.

The afterglow of GRB\,171205A was observed in multi-wavelength from near-ultraviolet to radio 
(e.g., Izzo et al. 2017; Butler et al. 2017; Mao et al. 2017a,b; de Ugarte Postigo et al. 2017a; Choi et al. 2017;
Melandri et al. 2017; Ramsay et al. 2017; Cobb et al. 2017; Smith \& Tanvir 2017; Laskar et al. 2017; Perley et al 2017).
The spectroscopic observation performed by VLT at 1.5hr after the trigger detected an 
optical transient at the position R.A.(J2000) = $11^{\mathrm h}09^{\mathrm m}39.573\mathrm{s}$ and Dec(J2000) =$-12\degr35\arcmin17.37\arcsec$, 
which is at the outskirts of the galaxy 2MASX\,J11093966-1235116 with a redshift of 0.037. The spectrum shows evident 
H$\alpha$, [\ion{N}{2}]$\lambda6583$ and [\ion{S}{2}]$\lambda\lambda$6716,6731 emission lines at the same redshift of the 
galaxy. 
The detection of an associated SN (SN\,2017iuk) was reported in December 07, 2017 by a 
follow-up spectroscopic observation using the 10.4m GTC telescope 
(de Ugarte Postigo et al. 2017b). The SN spectrum at that epoch is reported to be similar to the very early spectra of SN\,1998bw, which 
is further confirmed in Prentice et al. (2017).

\section{Observations and Data Reductions} \label{subsec:tables}

Our spectroscopic observations and date reductions are described in this section.

\subsection{Observations}

The long-slit spectra of GRB\,171205A/SN\,2017iuk were obtained by the NAOC
2.16m telescope at Xinglong observatory (Fan et al. 2016) at five epochs, i.e., December 17, 21, 25, 28 and 30. 
Figure 1 shows the R-band image of the field of GRB\,171205A taken by the 2.16m telescope at December 28, 2017. 
Our spectroscopic observations were carried out with the Beijing Faint Object Spectrograph and Camera (BFOSC) 
equipped with a back-illuminated E2V55-30 AIMO CCD as the detector. The grating G4 and a slit of width of 1.8\arcsec\
oriented in the south-north direction were used in all the five observation runs. This setup finally results 
in a spectral resolution of $\sim10$\AA\ as measured from the sky emission lines and comparison arcs, and provides
a wavelength coverage from 3850\AA\ to 8000\AA.
Except for the observation run at December 21, 2017, the target was observed twice in succession in each run.
The exposure time of each frame ranges from 1800 to 4800s, depending on both brightness of the object 
and weather condition. In each run, the wavelength calibration and flux calibration 
were carried out by the iron-argon comparison arcs and by the Kitt Peak National Observatory (KPNO) standard 
stars (Massey et al. 1988), respectively. The spectra of the standard stars were observed with the same instrumental 
setups immediately after the exposure of the object.
 
\begin{figure}[ht!]
\plotone{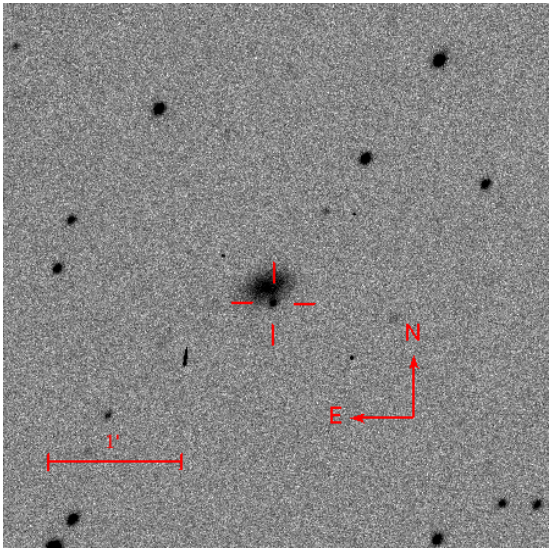}
\caption{R-band image of GRB\,171205A/SN\,2017iuk and its host galaxy 2MASX\,J11093966-1235116, which was taken at December 28, 2017, i.e.,
+23d after the trigger of the GRB. North is up, and east the left. The angular scale of the image is shown at the 
left-bottom corner. The burst, marked by the red cross, occurs at the outskirts of 
the host.  }
\end{figure}

\subsection{Data reductions}

Standard procedures were adopted to reduce the two-dimensional spectra 
by using the IRAF package\footnote{IRAF is distributed by the National Optical Astronomical Observatories,
which is operated by the Association of Universities for Research in
Astronomy, Inc., under cooperative agreement with the National Science
Foundation.}. The data reduction includes bias subtraction,
flat-field correction, and image combination along with cosmic-ray removal before the
extraction of the one-dimensional spectra\footnote{The image combination is skipped for the spectrum taken at 
December 21, 2017 since there was only one exposure. The cosmicray removal was performed on the single exposure before spectral extraction.}. 
The contamination due to the underlying host galaxy was taken into account in our extraction through background subtraction.
In order to reproduce the gradient of the surface brightness profile of the host,
the level of the underlying background is determined by a linear fitting in the two selected background regions. \rm
All the extracted one-dimensional spectra were then calibrated in wavelength and flux by the corresponding
comparison arc and standards. The flux calibration was performed by comparing the observed spectra of the standards with 
the spectrophtometrically calibrated spectra provided in the IRAF package, which corrects the specific response of both telescope and
spectrograph and the extinction due to Earth's atmosphere.
The A-band telluric feature around $\lambda\lambda$7600-7630 due to 
$\mathrm{O_2}$ molecules was removed from each observed spectrum by the corresponding standard.
The Galactic extinction was corrected by the extinction magnitude of $A_\mathrm{V}=0.138$ (Schlafly \& Finkbeiner 2011) 
taken from the NASA/IAPC Extragalactic Database (NED), assuming the $R_\mathrm{V} = 3.1$ extinction law of our
Galaxy (Cardelli et al. 1989). 
The spectra were then transformed to the rest frame, along with the correction of the relativity effect
on the flux, according to the redshift of 0.037 of the host galaxy.

\section{Analysis and Results}


\subsection{Identification and evolution}

Figure 2 shows the spectral evolution of SN\,2017iuk in the period from +12 to +30day. 
The first spectrum taken at +12 days after the onset of the GRB is very blue and featureless, 
except for the notch at $\sim6000$\AA\ caused by the
\ion{Si}{2}$\lambda6355$ absorption feature and the
two peaks around 4500\AA\ and 5300\AA. The latter two features are resulted from the \ion{Fe}{2}$\lambda5169$ absorption
(e.g., Filippenko 1997). All these features are quite typical for other GRB-SN events (e.g., Cano et al. 2017; 
D'Elia et al. 2015), which allows us to classify SN\,2017iuk as a SN IcBL with a highly stripped progenitor. \rm


One can see from Figure 2 that the two peaks at $\sim4500$\AA\ and $\sim5300$\AA\ gradually weaken from +12 to 23day, along with a gradual redshift for
the $\sim5300$\AA\ feature. The bottom spectrum shows that the emission from SN\,2017iuk fades out at +30day, in which 
there are marginal $\sim5300$\AA\ feature and 
extremely weak [\ion{O}{1}]$\lambda\lambda$6300, 6363 broad emission that is commonly detected in the nebular phase (e.g., Filippenko 1997).

\begin{figure}
\plotone{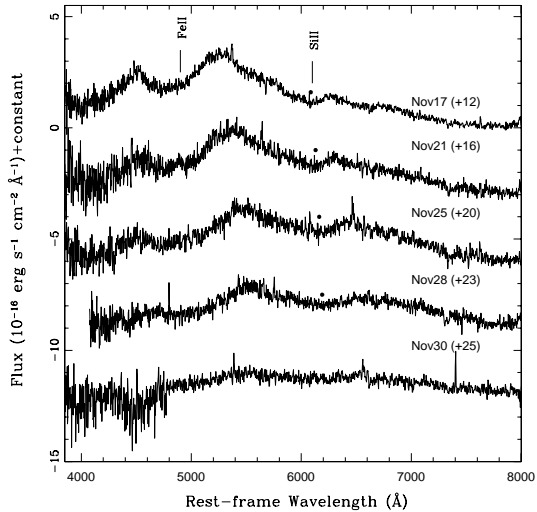}
\caption{Spectroscopic time-series of type IcBL SN\,2017iuk at the five different epochs from December 17 to 30, 2017.
All the spectra are transformed to the rest frame based on the redshift of 0.037, and are shifted vertically by an arbitrary amount 
for visibility. The \ion{Fe}{2}$\lambda5169$ and 
\ion{Si}{2}$\lambda$6355 absorption features are marked on the first spectrum. The black dots show the evolution of the position 
of the \ion{Si}{2}$\lambda$6355 absorption.}
\end{figure}

Figure 3 compares the spectrum of SN\,2017iuk taken at December 17, 2017 (+12day), which is close to the 
R-band light peak (Wang, X. G. et al. private communication)\rm, to the spectra of SN\,1998bw and SN\,2006aj at the similar epochs. Our comparison clearly 
suggests an analogy between SN\,2017iuk and SN\,2006aj.

\begin{figure}[ht!]
\plotone{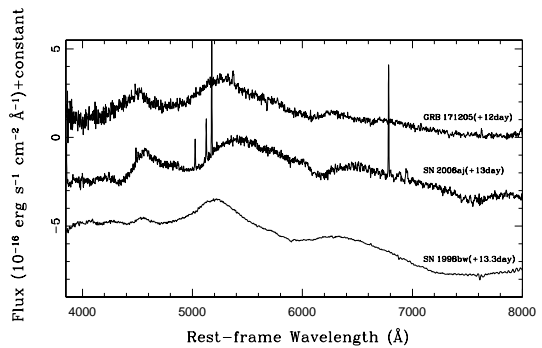}
\caption{A comparison of the spectrum of SN\,2017iuk taken at December 17, 2017 (+12day)
to those of SN\,1998bw (+13.3day) and SN\,2006aj (+13.0day) at the similar epochs.}
\end{figure}

\subsection{Photospheric velocity}

The photospheric expansion velocity in Type Ibc SNe is traditionally estimated from the \ion{Fe}{2}
lines at $\sim$5000\AA, because compared to the \ion{Fe}{2} lines the other ones are produced far above 
the photosphere. However, this method is unavailable for SNe IcBL due to their high 
velocities that result in a line identification difficulty because of the heavy line blending.   

We here attempt to estimate photospheric velocity of SN\,2017iuk by using the 
absorption trough of the \ion{Si}{2}$\lambda$6355 line (e.g., Sahu et al. 2018), except for the last spectrum taken at December 30, 2017.
In the last spectrum, the \ion{Si}{2}$\lambda$6355 feature is too weak to be measured. 
We mark the positions of the \ion{Si}{2}$\lambda$6355 line by dots in Figure 2, which clearly shows an evolution of 
the expansion of photosphere with a gradually decreasing velocity. The \ion{Si}{2}-based photospheric velocities (see Column (2) in Table 1)
decrease from 12,000 to 8,000$\mathrm{km\ s^{-1}}$ in the period from +12 to +23day. 
This temporal evolution of photospheric velocity is compared to the photospheric velocities of other type Ic SNe measured from the \ion{Si}{2} line
in Figure 4. 
One can again identify an analogy to SN\,2006aj from the evolution of photospheric velocity.

\begin{figure}[ht!]
\plotone{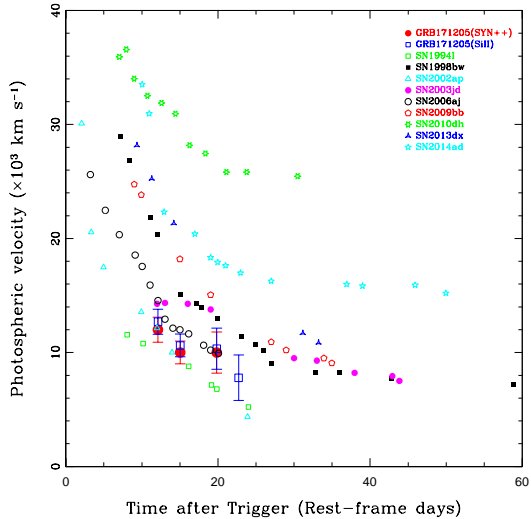}
\caption{Temporal evolution of the photospheric velocity of SN\,2017iuk. 
The blue open squares mark the photospheric velocities measured from the \ion{Si}{2}$\lambda6355$
absorption feature, and the red solid circles the velocities obtained from the fitting based on the
synthetic spectra generated by SYN++. The temporal evolutions of photospheric velocities of other type Ic SNe 
measured from the \ion{Si}{2}$\lambda$6355 features are 
plotted for a comparison. The data are taken from Sahu et al. (2018).}
\end{figure}

\begin{table}[h!]
\renewcommand{\thetable}{\arabic{table}}
\centering
\caption{Photospheric velocities and blackbody temperatures measured from the spectra} \label{tab:decimal}
\begin{tabular}{cccc}
\tablewidth{0pt}
\hline
\hline
Date & $\upsilon_{\mathrm{ph}}(\mathrm{SiII})$ & $\upsilon_{\mathrm{ph}}(\mathrm{syn})$ & $T_{\mathrm{bb}}$\\ 
     & $\mathrm{km\ s^{-1}}$ & $\mathrm{km\ s^{-1}}$ & K \\
(1)  &   (2) & (3)  & (4)\\
\hline
Dec 17, 2017 & $12700\pm1100$ & $12000\pm1000$ &  12000\\
Dec 21, 2017 & $10600\pm1000$ & $10000\pm1000$  &   8000\\
Dec 25, 2017 & $10300\pm1800$ & $10000\pm2000$  &   6000\\
Dec 28, 2017 & $7800\pm2000$  & \dotfill         &  \dotfill\\
\hline
\hline
\end{tabular}
\end{table}

\subsection{Spectral modeling with SYN++}

In this section, we model the spectra of the SN\,2017iuk through the 
synthetic spectra generated by the SYN++ code (Thomas et al. 2011) that is an enhanced 
version of the parameterized supernova spectrum synthesis code SYNOW (Fisher 2000; Branch et al. 2000).
In generating synthetic spectra, the exciting temperature is fixed to be 6,000K, and the involved 
ions include \ion{Fe}{2}, \ion{Co}{2}, \ion{Si}{2}, \ion{Ca}{2}, \ion{Mg}{2} and \ion{O}{1}.
Based on the synthetic spectra, we fit the observed rest-frame spectra over the whole spectroscopic wavelength range
through ``chi-by-eye'' by changing the blackbody temperature, the optical depth of each ion, 
and the velocity of the photosphere. The fittings are schemed in Figure 5 for the spectra taken at 
December 17, 21 and 25, 2017.  One can see from the figure that in all the three cases the generated
synthetic spectra  generally match the observed ones quite well, except for the ``blue wing'' of the 
$\sim5000$\AA\ feature. 
This failure of reproducing might be due to the imperfect of the adopted atomic data of iron, 
especially when the photospheric temperature is low. \rm
The best-fitted photospheric velocities decrease from 12,000 to 10,000$\mathrm{km\ s^{-1}}$ within
the period from +12 to +20d after the trigger of the GRB. The best-fitted velocities are overplotted in 
Figure 4, which shows a significant consistence with the measurements based on 
the \ion{Si}{2} absorption.
In the period, the modeled blackbody temperatures decrease from 12,000 to
6000K. The modeled photospheric velocities and blackbody temperatures are tabulated in Table 1 (columns (3) and (4)).

\begin{figure}[ht!]
\plotone{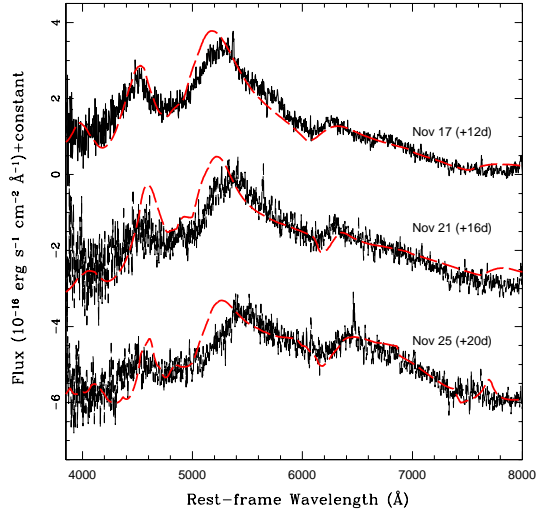}
\caption{An illustration of the spectral fitting based on the synthetic spectra generated by using 
SYN++.}
\end{figure}



\section{Conclusion and Discussion} \label{sec:pubcharge}

We monitored SN\,2017iuk associated with the \it ll\rm GRB GRB\,171205A in 
spectroscopy, which is up to now the third GRB-SN event away from us.
Our spectroscopy observations and spectral analysis enable us to identify the SN as a broad-line type Ic SN.
A comparison study suggests that SN\,2017iuk resembles to SN\,2006aj due to their 1) similar spectra at the similar epochs and 2) consistent evolutions of 
photospheric velocity.

\subsection{Explosion parameters and mechanism}

In this section, we estimate the explosion parameters from our spectral analysis
presented in Section 4, and argue that the estimated parameters roughly agree with those of 
SN\,2006aj, which reinforces the analogy revealed in the above section.

%

We estimate the mass of ejecta $M_{\mathrm{ej}}$ from the evolution of photospheric velocity 
by assuming an exponential density profile $\rho=\rho_0 e^{-\upsilon/\upsilon_e}$, where 
$\rho_0$ is the central density and $\upsilon_e$ the e-folding velocity. The evolution
of photospheric velocity can be therefore expressed as (Eq. 3 in Deng et al. 2001) 
\begin{equation} 
 \upsilon_{\mathrm{ph}}\approx -\upsilon_e \ln\bigg[10^{-6}\frac{\tau_{\mathrm{ph}}}{\overline{\kappa}}
 \bigg(\frac{\upsilon_e}{10^3\mathrm{km\ s^{-1}}}\bigg)^2\bigg(\frac{M_{\mathrm{ej}}}{M_\odot}\bigg)^{-1}\bigg]
 -2\upsilon_e\ln t_{\mathrm{d}}
\end{equation}
where $\tau_{\mathrm{ph}}$ is the optical depth at the photosphere, $\overline{\kappa}$ the optical opacity
and $t_{\mathrm{d}}$ the time since explosion in unit of day. 
With the measured $\upsilon_{\mathrm{ph}}$, we fitted the evolution of $\upsilon_{\mathrm{ph}}$ as 
$\upsilon_{\mathrm{ph}}=a+b\ln t_{\mathrm{d}}$. After deriving a value of $\upsilon_e$ from the 
best fitted $b$ ($=-2\upsilon_e$), the ejecta mass is inferred to be $M_{\mathrm{ej}}\approx 1.1M_\odot$ from the 
best fitted value of $a$, where the typical values of $\tau_{\mathrm{ph}}=1$ and $\overline{\kappa}=0.07\mathrm{cm^2\ g^{-1}}$
are adopted in the estimation.
The explosion kinetic energy is then estimated to 
$E_{\mathrm{k}}=6M_{\mathrm{ej}}\upsilon_e^2=1.4\times10^{51}\ \mathrm{erg}$ by integrations of both $\rho$
and $\rho\upsilon^2$ over velocity. 
Although the inferred $M_{\mathrm{ej}}$ and $E_{\mathrm{k}}$ are within the ranges of the typical values of SNe IcBL, they are
at the lower end of the distributions of the GRB-associated SNe IcBL that have the typical 
values of $M_{\mathrm{ej}}\sim 1-10M_\odot$ and $E_{\mathrm{k}}\sim1\times10^{52}\ \mathrm{erg}$
(e.g., Cano et al. 2017).

We alternatively estimate $E_{\mathrm{k}}$ by using the expression of $E_{\mathrm{k}}=3/10M_{\mathrm{ej}}\upsilon_{\mathrm{ph}}^2$
given in Arnett (1982, 1996). The photospheric velocity at the time of bolometric maximum, i.e., +12d after the trigger,
results in a value of $E_{\mathrm{k}}\approx9\times10^{50}\ \mathrm{erg}$, which is roughly consistent with the above
value that is estimated from $\upsilon_e$.
The $E_{\mathrm{k}}/M_{\mathrm{ej}}$ ratio is resulted to be $\sim0.13$ for SN\,2017iuk.

We argue that the explosion parameters estimated above are comparable to those of SN\,2006aj (see Table 3 in Cano et al. 2017).
The comparable explosion parameters and the revealed similarity in the spectral evolution 
motivate us to suspect that SN\,2017iuk/GRB\,171205A is of the similar explosion mechanism to SN\,2006aj/GRB,060218.
By a detailed modeling of the light curve and spectra, Mazzali et al. (2006) suggested \rm that SN\,2006aj/GRB\,060218
is produced by a core-collapsing of a massive star with an initial mass of $\sim20M_\odot$, which expects a 
formation of a neutron star rather than a black hole after the core-collapse.

\subsection{Host galaxy}

The host galaxy 2MASX\,J11093966-1235116 (LCRS\,B110709.2-121854) of GRB\,171205A/SN\,2017iuk is classified as a S0 galaxy in the
Lyon Extragalactic Database (LEDA). The poor seeing (2.5-3\arcsec) of our observations, however, prevent us from 
further morphological study on the galaxy. We estimate the total stellar mass ($M_\star$) of the galaxy from its K-band photometry,
because the near-infrared emission traces the mass of late-type stars better and is much less sensitive to 
extinction by dust. With the distance modulus of $\mu=35.90\pm0.15$mag, the absolute magnitude in $K_s$-band is obtained to be 
$-23.57\pm0.21$mag, which yields a luminosity in $K_s$-band of $L_{K}=5.5\times10^{10}L_\odot$ by adopting 
an absolute solar $K_s$ band magnitude of 3.29 (Blanton \& Roweis 2007). Adopting a universal $K_s$-band 
mass-to-light ratio $M/L=0.6$ (Bell \& de Jong 2000) finally returns a $M_\star=3.3\times10^{10}M_\odot$, which
is above the average stellar mass of the hosts of nearby LGRBs (e.g., Kruhler et al. 2015; Perley et al. 2016; Schulze et al. 2015)

The spectrum of the nucleus of the host galaxy has been taken by the 6dF Galaxy Survey (Jones et al. 2004),
which is a spectroscopic survey using the UK Schmidt telescope at Anglo-Australian Observatory. 
The using of robotic positioning optical fibers allows the telescope to measure distances for more than 100,000 galaxies in 6 years.  
The rest-frame spectrum\footnote{The rest-frame spectrum is transformed from the 
observed one by applying both Galactic extinction and Doppler corrections. See section 3.2.2 for the description in details.} 
extracted from the final data release (DR3, Jones et al. 2009) is shown in Figure 6. 
It is noted that the spectrum shown in the figure is lack of absolute flux calibration.   
Nevertheless, the spectrum clearly shows that the host galaxy of GRB\,171205A is a typical starforming galaxy 
with strong H$\alpha$, H$\beta$, [\ion{N}{2}]$\lambda\lambda$6548, 6583, [\ion{S}{2}]$\lambda\lambda$6727,6731, and 
weak [\ion{O}{3}]$\lambda\lambda$4959, 5007 emission lines. 
Figure 7 shows the occupation on the 
two empirical Baldwin-Phillips-Terlevich (BPT) diagrams for the host galaxy. The diagrams, which were 
originally proposed by Baldwin et al. (1981), and then refined by Veilleux \& Osterbrock (1987), 
are traditionally used as a powerful tool to determine the dominant
energy source in emission-line galaxies according to their emission-line ratios.
By measuring the flux of each emission line through direct integration, the figure shows that the host galaxy is located within
the locus of starforming sequence very well. 
\begin{figure}[ht!]
\plotone{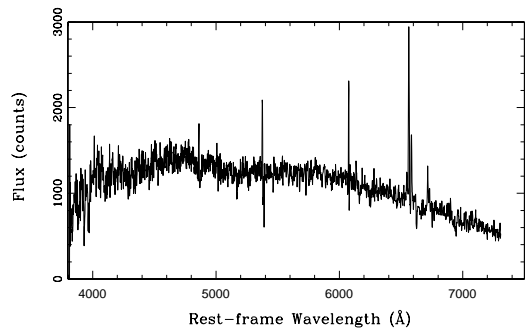}
\caption{Rest-frame spectrum of the nucleus of the host galaxy of GRB\,171205A/SN\,2017iuk taken by the 6dF Galaxy Survey. 
Note that the spectrum is lack of absolute flux calibration. }
\end{figure}

\begin{figure}[ht!]
\plotone{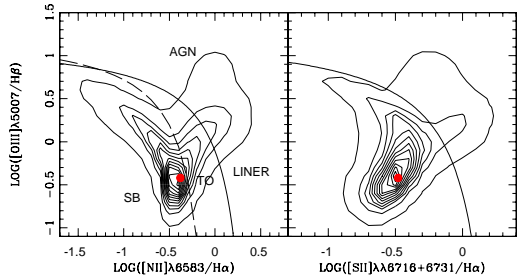}
\caption{Two BPT diagnostic diagrams for the host galaxy of GRB\,171205A/SN\,2017iuk. 
The location of the burst is marked by the red circle in
both panels, where the density contours are shown for a typical
distribution of the narrow-line galaxies described in Heckman et al. (2004) and Kauffmann et al. (2003). 
Only the galaxies with S/N$>$20 and the emission lines detected with
at least 3$\sigma$ significance are plotted. The solid lines in both panels mark the theoretical
demarcation lines separating AGNs from star-forming galaxies proposed by Kewley et al. (2001). The long-dashed line 
in the left panel shows the empirical demarcation line proposed in
Kauffmann et al. (2003), which is used to separate ``pure'' star-forming galaxies. }
\end{figure}

We then estimate the star formation rate (SFR) from its near-ultraviolet detection. The brightness at the 
NUV band (peaked at $\sim2800$\AA) of GALEX is reported to be $18.12\pm0.04$mag. At the beginning, the NUV
brightness is corrected for the Milkyway dust extinction from color excess through $A_{\mathrm{NUV}}=R_{\mathrm{NUV}}E(B-V)$.
The parameter $R_{\mathrm{NUV}}$ is determined to be 6.62 (Fitzpatrick 1999), and updated to be $7.24\pm0.08$ by
Yuan et al. (2013). The calibration of $\mathrm{SFR}=1\times10^{-28}L_{\mathrm{NUV}}\ \mathrm{M_\odot\ yr^{-1}}$,
obtained by converting the relation from Kennicutt (1998)
to the IMF in Kroupa (2001), is used to estimate SFR, where $L_{\mathrm{NUV}}$ is the NUV luminosity in unit of 
$\mathrm{erg\ s^{-1}}$. With the extinction-corrected NUV brightness, we finally obtain a SFR of $\sim1\mathrm{M_\odot\ yr^{-1}}$
for the host galaxy of GRB\,171205A/SN\,2017iuk, which is lager than the average value of the host galaxies of LGRBs at the 
similar redshift (e.g., Kruhler et al. 2015). With the estimated stellar mass of the host,
the specific SFR (sSFR), defined as 
the SFR normalized to the total stellar mass, is inferred to be as low as $\sim0.03\mathrm{Gyr^{-1}}$. In fact, 
this value is at the lower end of the distribution of sSFR of samples of LGRB hosts (e.g. Savaglio et al. 2009; Japelj et al. 2016).  
Assuming a constant SFR over the growth history of the host, the growth timescale 
$t_\star=M_\star/\mathrm{SFR}$ is estimated to be about 30Gyr. This timescale is comparable to (or larger than) the Hubble time of the local universe,
which implies a quiescent growth of the host galaxy of SN\,2017iuk/GRB\,171205A.  
 
We finally estimate nuclear metallicity of the host galaxy of GRB\,171205A/SN\,2017iuk from the spectrum. 
The oxygen metallicity is simply calculated from the $N2$ method proposed in Pettni \& Pagel (2004): 
$\mathrm{12+\log(O/H)=8.90+0.57}N2$, where $N2=\mathrm{\log([NII]/H\alpha)}$, because the [\ion{N}{2}]/H$\alpha$
line ratio is insensitive to both flux calibration and intrinsic dust extinction. The metallicity of the galaxy is inferred
to be $\mathrm{12+\log(O/H)}=8.69$, which equals to the solar gas phase value (Allende Prieto et al. 2001; Asplund et al. 2004). 
In fact, by using strong-line diagnostic, Kruhler et al. (2015) reported that the oxygen metallicities range from 7.0 to 
9.0 for a sample of 44 LGRBs within a redshift range from 0.3 to 3.4.
 
In summary, GRB\,171205A/SN\,2017iuk occurred in an early type, high-mass, starforming galaxy with
low sSFR and solar metallicity.

\acknowledgments

The authors thank the anonymous referee for a careful
review and helpful suggestions that improved the manuscript.
The study is supported by the National Basic Research
Program of China (grant 2014CB845800) and by the Strategic
Pioneer Program on Space Science, Chinese Academy of
Sciences, Grant No.XDA15052600. 
JW is supported by the National Natural Science Foundation of China under grants
11473036 and 11773036. 
DX acknowledges the supports by the One-Hundred-Talent Program of the 
Chinese Academy of Sciences (CAS), by the Strategic Priority Research 
Program “Multi-wavelength Gravitational Wave Universe” of the CAS (No. 
XDB23000000), and by the National Natural Science Foundation of China 
under grant 11533003.
Special thanks go to the staff at Xinglong Observatory
as a part of National Astronomical Observatories, China
Academy of Sciences for their instrumental and observational help, and 
to the allocated observers who allow us to finish the observations in ToO mode.
This study is partially supported by the Open Project Program of
the Key Laboratory of Optical Astronomy, NAOC, CAS. The study uses the data collected by 6dF Galaxy Survey which was
carried out by the staff of the Australian Astronomical Observatory.

\vspace{5mm}
\facilities{Xinglong Observatory 2.16m telescope}
\software{IRAF (Tody 1986, 1993), SYN++ (Thomas et al. 2011), SYNOW (Fisher 2000; Branch et al. 2000)}


\begin{thebibliography}{}

\bibitem[Allende Prieto et al. (2001)]{all01} Allende Prieto, C., Barklem, P. S., Asplund, M., \& Ruiz Cobo, B. 2001, \apj, 558, 830
\bibitem[Arnett (1982)]{arn82} Arnett, W. D. 1982, \apj, 253, 785
\bibitem[Arnett (1996)]{arn96} Arnett, W. D. 1996, Supernovae and Nucleosynthesis: An Investigation of the History of Matter from the Big Bang to the 
Present, Princeton University Press
\bibitem[Asplund et al. (2004)]{asp04} Asplund, M., Grevesse, N., Sauval, A. J., Allende Prieto, C., \& Kiselman, D. 2004, \aap, 417, 751
\bibitem[Baldwin et al. (1981)]{bal81} Baldwin, J. A., Phillips, M. M., \& Terlevich, R. 1981, \pasp, 93, 5
\bibitem[Barthelmy et al. (2017)]{bar17} Barthelmy, S. D., Cummings, J. R., D'Elia, V., et al. 2017, GCN, 22184, 1
\bibitem[Bersten et al. (2016)]{ber16} Bersten, M. C., Benvenuto, O. G., Orellana, M., \& Nomoto, K. 2016, \apjl, 817, 8 
\bibitem[Bertin \& Arnouts (1996)]{bea96} Bertin, E., \& Arnouts, S. 1996, \aaps, 117, 393
\bibitem[Bell \& de Jong (2000)]{bed00} Bell, E. F., \& de Jong, R. S. 2000, \mnras, 312, 497
\bibitem[Blondin \& Tonry (2007)]{blt07} Blondin, S., \& Tonry, J. L. 2007, \apj, 666, 1024
\bibitem[Branch etal. (2000)]{bra00} Branch, D., Jeffery, D. J., Blaylock, M., \& Hatano, K. 2000, \pasp, 112, 217
\bibitem[Bucciantini et al. (2007)]{buc07} Bucciantini, N., Quataert, E., Arons, J., Metzger, B. D., \& Thompson, T. A. 2007, \mnras, 380, 1541
\bibitem[Butler et al. (2017)]{but17} Butler, N., Watson, A. M., Kutyrev, A., et al. 2017, GCN, 22182, 1 
\bibitem[Cano (2014)]{can14} Cano, Z. 2014, \apj, 794, 121
\bibitem[Cano et al. (2017)]{can17} Cano, Z., Wang, S. Q., Dai, Z. G., \& Wu, X. 2017, AdAst, 2017, 5
\bibitem[Cano et al. (2016)]{can17} Cano, Z., Johansson A. K. G., \& Maeda, K. et al. 2016, \mnras, 457, 2761
\bibitem[Cardelli et al. (1989)]{car89} Cardelli, J. A., Clayton, G. C., \& Mathis, J. S. 1989, \apj, 345, 245
\bibitem[Choi et al. (2017)]{cho17} Choi, C., Im, M., Gak, L. S., \& Pan-Starrs, et al. 2017, GCN, 22188, 1
\bibitem[Cobb (2017)]{cob17} Cobb, B. E. 2017, GCN, 22270, 1
\bibitem[Dai \& Lu (1998)]{dai98} Dai, Z. G., \& Lu, T. 1998, PhRvL, 81, 4301
\bibitem[Dai et al. (2016)]{dai16} Dai, Z. G., Wang, S. Q., Wang, J. S., Wang, L. J., \& Yu, Y. W. 2016, \apj, 817, 132
\bibitem[D'Elia et al. (2015)]{del15} D'Elia, V., Pian, E., Melandri, A., et al. 2015, \aap, 577, 116 
\bibitem[D'Elia et al. (2017a)]{del17a} D'Elia, V., D'Ai, A., Lien, A. Y., \& Sbarufatti, B. 2017, GCN, 22177, 1
\bibitem[D'Elia et al. (2017b)]{del17b}	D'Elia, V., D'Ai, A., Melandri, A., et al. 2017, GCN, 22271, 1
\bibitem[de Jong et al. (2004)]{dej04} de Jong, R. S., Simard, L., Davies, R. L., et al. 2004, \mnras, 355, 1155
\bibitem[Della Valle et al. (2006)]{del06} Della Valle, M., Chincarini, G., Panagia, N., et al. 2006, \nat, 444, 1050 
\bibitem[de Ugarte Postigo et al. (2017a)]{deu17a} de Ugarte Postigo, A., Schulze, S., Bremer, M., et al. 2017a, GCN, 22187, 1 
\bibitem[de Ugarte Postigo et al. (2017b)]{deu17b} de Ugarte Postigo, A., Izzo, L., Kann, D. A., Thoene, C. C., Pesev, P., 
Scarpa, R., \& Perez, D. 2017b,  ATel, 11038, 1
\bibitem[Deng et al (2001)]{den01} Deng, J., Qiu, Y. L., \& Hu, J. Y. 2001, arXiv:astro-ph/0106404
\bibitem[Fan et al. (2016)]{fan16} Fan, Z., Wang, H. J., Jiang, X. J., et al., \pasp, 128, 5005
\bibitem[Filippenko (1997)]{fil97} Filippenko, A. V. 1997, \araa, 35, 309
\bibitem[Fisher (2000)]{fis00} Fisher, A. K. 2000, PhD thesis, Univ. Oklahoma
\bibitem[Fitzpatrick (1999)]{fit99} Fitzpatrick E. L., 1999, \pasp, 111, 63
\bibitem[Frederiks et al. (2017)]{fre17} Frederiks, D., Golenetskii, S., Aptekar, R., et al. 2017, GCN, 22227, 1
\bibitem[Fynbo et al. (2006)]{fyn06} Fynbo, J. P. U., Watson, D., Thone, C. C., et al. 2006, \nat, 444, 1047
\bibitem[Galama et al. (1998)]{gal98} Galama, T. J., Vreeswijk, P. M., van Paradijs, J., et al. 1998, \nat, 395, 670
\bibitem[Gal-Yam et al. (2006)]{gal06} Gal-Yam, A., Fox, D. B., Price, P. A., et al. 2006, \nat, 444, 1053
\bibitem[Greiner et al. (2015)]{gre15} Greiner, J. Mazzali, P. A., Kann, D., et al. 2015, \nat, 523, 189
\bibitem[Heckman et al. (2004)]{kec04} Heckman, T. M., Kauffmann, G., Brinchmann, J., Charlot, S., Tremonti, C., \&
White, S. D. M. 2004, \apj, 613, 109
\bibitem[Hjorth \& Bloom (2012)]{hjb12} Hjorth, J., \& Bloom, J. S. 2012, in Gamma-Ray Bursts, Vol. 51, ed.
C. Kouveliotou, R. A. M. J. Wijers, \& S. Woosley (Cambridge: Cambridge Univ. Press), 169
\bibitem[Iwamoto et al. (2000)]{iwa00} Iwamoto, K., Nakamura, T., Nomoto, K., et al. 2000, \apj, 534, 660
\bibitem[Izzo et al. (2017)]{izz17} Izzo, L., Selsing, J., Japelj, J., et al. 2017, GCN, 22180, 1
\bibitem[Japelj et al. (2016)]{jap16} Japelj, J., Vergani, S. D., Salvaterra, R. et al. 2016, \aap, 590, 129 
\bibitem[Jones et al. (2009)]{jon09} Jones, D. H., Read, M. A., Saunders, W., et al. 2009, \mnras, 399, 683
\bibitem[Jones et al. (2004)]{jon04} Jones, D. H., Saunders, W., Colless, M., et al. 2004, \mnras, 355, 747
\bibitem[Kasen \& Bildsten (2010)]{kab10} Kasen, D., \& Bildsten, L. 2010, \apj, 717, 245
\bibitem[Kauffmann et al. (2003)]{kau03} Kauffmann, G., et al. 2003, \mnras, 346, 1055
\bibitem[Kennea et al. (2017)]{ken17} Kennea, J. A., Sbarufatti, B., Burrows, D. N., et al. 2017, GCN, 22183, 1
\bibitem[Kennicutt (1998)]{ken98} Kennicutt, R. C. Jr. 1998, \araa, 36, 189
\bibitem[Kewley et al. (2001)]{kew01} Kewley, L. J., Dopita, M. A., Sutherland, R. S., Heisler, C. A., \& Trevena, J. 2001, \apj, 556, 121
\bibitem[Kroupa (2001)]{kru01} 	Kroupa, P. 2001, \mnras, 322, 231
\bibitem[Kruhler et al. (2015)]{kru15} Kruhler, T., Malesani, D., Fynbo, J. P. U., et al. 2015, \aap, 581, 125
\bibitem[Laskar et al. (2017)]{las17} Laskar, T., Coppejans, D. L., Margutti, R., \& Alexander, K. D. 2017, GCN, 22216, 1
\bibitem[Lazzati et al. (2001)]{laz01}Lazzati, D., Covino, S., Ghisellini, G., et al. 2001, \aap, 378, 996
\bibitem[MacFadyen \& Woosley (1999)]{maw93} MacFadyen, A. I., \& Woosley, S. E. 1999, \apj, 524, 262
\bibitem[Maeda et al. (2003)]{mae03} Maeda, K., Mazzali, P. A., Deng, J. S., et al. 2003, \apj, 593, 931
\bibitem[Mao et al. (2017a)]{mao17a} Mao, J., Ding, X., \& Bai, J. -M. 2017a, GCN, 22186, 1
\bibitem[Mao et al. (2017b)]{mao17b} Mao, J., Ding, X., \& Bai, J. -M. 2017b, GCN, 22195, 1
\bibitem[Masetti et al. (2005)]{mas05} Masetti, N., Palazzi, E., Pian, E., et al. 2005, \aap, 438, 841
\bibitem[Massey et al. (1988)]{mas88} Massey, P., Strobel, K., Barnes, J. V., et al. 1988, \apj, 328, 315
\bibitem[Mazzali et al. (2006)]{maz06} Mazzali, P. A., Deng, J., Nomoto, K., et al. 2006, \nat, 442, 1018 
\bibitem[Mazzali et al. (2014)]{maz14} Mazzali, P. A., McFadyen, A. I., Woosley, S. E., Pian, E., \& Tanaka, M. 2014, \mnras, 443, 67
\bibitem[Melandri et al. (2017)]{mel17} Melandri, A., D'Avanzo, P., di Fabrizio, L., Padilla, C., \& D'Elia, V. 2017, GCN, 22189, 1
\bibitem[Metzger et al. (2015)]{met15} Metzger, B. D., Margalit, B., Kasen, D., \& Quataert, E. 2015, \mnras, 454, 3311
\bibitem[Nakamura et al. (2001)]{nak01} Nakamura, T., Mazzali, P. A., Nomoto, K., \& Iwamoto, K. 2001, \apj, 550, 991
\bibitem[Perley et al. (2017)]{per17} Perley, D. A., Schulze, S., \& de Ugarte Postigo, A. 2017, GCN, 22252, 1
\bibitem[Perley et al. (2016)]{per16} Perley, D. A., Tanvir, N. R., Hjorth, J., et al. 2016, \apj, 817, 8 
\bibitem[Pian et al. (2006)]{pia06} Pian, E., Mazzali, P. A., Masetti, N., et al. 2006, \nat, 442, 1011
\bibitem[Prentice et al. (2017)]{pre17} Prentice, S., Mazzali, P., Smartt, S. J. et al. 2017, ATel, 11060, 1 
\bibitem[Sahu et al. (2018)]{shu18} Sahu, D. K., Anupama, G. C., Chakradhari, N. K., Srivastav, S., Tanaka, M., Maeda, K., \& 
Nomoto, K. 2018, \mnras, 475, 2591
\bibitem[Savaglio et al. (2009)]{sav09} Savaglio, S., Glazebrook, K., \& Le Borgne, D. 2009, \apj, 691, 182
\bibitem[Schlafly \& Finkbeiner (2011)]{scf11} Schlafly, E. F., \& Finkbeiner, D. P. 2011, \apj, 737, 103
\bibitem[Schulze et al. (2015)]{sch15} 	Schulze, S., Chapman, R., Hjorth, J., et al. 2015, \apj, 808, 73
\bibitem[Smith \& Tanvir (2017)]{smt17} Smith, I. A., \& Tanvir, N. R. 2017, GCN, 22242, 1
\bibitem[Sukhbold et al. (2016)]{suk16} Sukhbold, T., Ertl, T., Woosley, S. E., Brown, J. M., \& Janka, H.-T. 2016, \apj, 821, 38
\bibitem[Thomas etal. (2011)]{tho11} Thomas, R. C., Nugent, P. E., \& Meza, J. C. 2011, \pasp, 123, 237 
\bibitem[Tody (1986)]{tod86} Tody, D. 1986, Proc. SPIE, 627, 733
\bibitem[Tody (1993)]{tod93} Tody, D. 1993, in ASP Conf. Ser. 52, adass II, ed. R. J. Hanisch, R. J. V. Brissenden, \& J. Barnes (San Francisco, CA: ASP), 173
\bibitem[Veilleux \& Osterbrock (1987)]{veo87} Veilleux, S., \& Osterbrock, D. E. 1987, \apjs, 63, 295
\bibitem[Wang et al. (2017)]{wan17b} Wang, L. J., Yu, H., Liu, L. D., et al. 2017, \apj, 837, 128
\bibitem[Wang et al. (2018)]{wan18} Wang, J., Xin, L. P., Qiu, Y. L., Xu, D. W., \& Wei, J. Y. 2018, \apj, 855, 91
\bibitem[Wang et al. (2015)]{wan15} Wang, S. Q., Wang, L. J., Dai, Z. G., \& Wu, X. F. 2015, \apj, 799, 107
\bibitem[Woosley (1993)]{woo93} Woosley, S. E. 1993, \apj, 405, 273
\bibitem[Woosley \& Bloom (2006)]{wob06} Woosley, S. E., \& Bloom, J. S. 2006, \araa, 44, 507
\bibitem[Woosley \& Heger (2006)]{woh06} Woosley, S. E., \& Heger, A. 2006, \apj, 637, 914
\bibitem[Yuan et al. (2013)]{yua13} Yuan, H. B., Liu, X. W., \& Xiang, M. S. 2013, \mnras, 430, 2188
\bibitem[Zhang \& Dai (2010)]{zhd10} Zhang, B., \& Dai, Z. G. 2010, \apj, 718, 841
\end{thebibliography}
\end{document}